\documentclass[twocolumn,superscriptaddress,secnumarabic,aps,pra,nofootinbib,numerical]{revtex4}
\usepackage[english]{babel}
\usepackage{epstopdf}
\usepackage{pdfpages}
\usepackage{float}
\usepackage{enumerate}

\usepackage{amsmath} 
\usepackage{dsfont}
\usepackage{cancel}
\usepackage{amsthm} 
\usepackage{amssymb}	
\usepackage{graphicx} 
\usepackage{empheq}
\usepackage[absolute,overlay]{textpos}
\usepackage{xcolor}

\allowdisplaybreaks

\let\vaccent=\v 

\renewcommand{\v}[1]{\ensuremath{\mathbf{#1}}} 
 
\let\baraccent=\= 
\renewcommand{\=}[1]{\stackrel{#1}{=}} 

\DeclareMathOperator{\Tr}{Tr}

\begin{document}

\title{Sensing Floquet-Majorana fermions via heat transfer}
\date{\today}

\author{Paolo Molignini}
\affiliation{Institute for Theoretical Physics, ETH Z\"{u}rich, 8093 Zurich, Switzerland}
\author{Evert van Nieuwenburg}
\author{R. Chitra}
\affiliation{Institute for Theoretical Physics, ETH Z\"{u}rich, 8093 Zurich, Switzerland}

\begin{abstract}
Time periodic modulations of the transverse field in the closed  XY  spin-$\frac12$ chain generate a very rich dynamical phase diagram, with a hierarchy of $\mathbb{Z}_n$ topological phases characterized by differing numbers of Floquet-Majorana modes.
This rich phase diagram survives when the system is coupled to dissipative end reservoirs.
Circumventing the obstacle of preparing and measuring quasi-energy configurations endemic to Floquet-Majorana detection schemes, we show that stroboscopic heat transport and spin density are robust observables to detect both the dynamical phase transitions and Majorana modes in dissipative settings.
We find that the heat current provides very clear signatures of these Floquet topological phase transitions.
In particular, we observe that the derivative of the heat current, with respect to a control parameter, changes sign at the boundaries separating topological phases with differing non-zero numbers of Floquet Majorana modes.
We present a simple scheme to directly count the number of Floquet-Majorana modes in a phase  from the Fourier transform of the local spin density profile.
Our results are valid provided the anisotropies are not strong and can be easily implemented in quantum engineered systems.
\end{abstract}

\maketitle

\textit{Introduction} ---  Recent developments in quantum engineering~\cite{Eisert} offer remarkable possibilities to probe physics in strongly out-of-equilibrium regimes.
Particularly interesting  from this perspective are  periodically driven quantum systems also known as Floquet systems.
Floquet systems, on the one hand open up fundamental questions about non-equilibrium steady states~\cite{Lazarides} and on the other offer a rich toolbox to explore new dynamical phases of matter. 
Examples of the latter include dynamically induced superfluid-Mott insulating transitions in optical lattices~\cite{Eckardt}, coherent destruction of tunneling\cite{Grifoni,Haenggi}, as well as dynamical many-body phases of parametrically driven systems with no static counterparts~\cite{OdedChitra}.

Periodic driving also offers the intriguing possibility of dynamically generating exotic topological excitations in otherwise topologically trivial systems~\cite{Lindner, Kitagawa, Titum, Gedik}. 
One of the most well known topological excitations are zero energy Majorana modes which, for example, occur as localized edge modes in static Kitaev models~\cite{Kitaev2001}.  
The non-abelian braiding statistics of Majorana modes makes them promising candidates for topological quantum computation~\cite{Kitaev2, Liu}. 
However, direct observations of these Majorana modes in quantum wires are challenging because of their intrinsic weak charge coupling. 
Indirect observations based on spectroscopy or interferometric measurements in proximitized semiconductor nanowire devices~\cite{AliceaReview,KTLaw,Fidkowski,Fu-Kane2009,Mourik} or hybrid superconducting-quantum interference devices~\cite{Woerkom:2016aq} are still debated as the signals are hard to distinguish from the contributions of other processes like Andreev bound states and the Kondo effect~\cite{Mourik, Plugge:2016cy}. 

Recently, it was shown that a hierarchy of Floquet Majorana fermions (FMF) could be generated in an isolated spin-$\frac{1}{2}$ (fermionic) chain subject to a periodically varying magnetic field (chemical potential)~\cite{Sedrakyan:2011,Thakurathi,Liu}. 
These exotic emergent and out-of-equilibrium modes are dynamically generated in the Floquet quasi-energy spectrum but retain familiar topological characteristics, like winding numbers.  
Toy models of FMF's were shown to lead to novel sum rules for differential conductance evaluated over all the quasi-energies in driven topological insulators~\cite{Kundu,YantaoLi}.  
However, practical realizations of transport based schemes to evaluate these sum rules  are hampered by difficulties in preparing the system in the appropriate energy interval~\cite{Kitagawa}, garnering a clear knowledge of chemical potential bias~\cite{Farrell-Pereg-Barnea} and extracting all quasi-energies simultaneously~\cite{Gedik}. 
Proposals for clear and universal signatures of Majorana modes, both in- and out-of-equilibrium are  thus very desirable.

In this Letter, we study a driven dissipative spin chain where a hierarchy of Floquet Majorana excitations and associated topological phase transitions can be induced in a controlled manner\cite{Thakurathi}. 
The possibility to easily tune in and out of different topological phases makes them ideal systems  for measuring exotic excitations. 
We construct stroboscopic observables allowing us not only to distinguish phases with different FMF's but also to count their number.
Our work offers a direct generalization of dynamically generated topology to more realistic open systems and  obviates the need  for special  initial state preparations  and fine tuning. 
We show that the stroboscopic heat current in this dissipative setup provides very clear signatures of  the cascade of transitions between topological  $\mathbb{Z}_n$ phases with differing numbers of FMF's.  
Moreover, the number of FMF's can be directly obtained from the stroboscopic spin density.  
\begin{center}
	\begin{figure}[H]
		\includegraphics[width=\columnwidth]{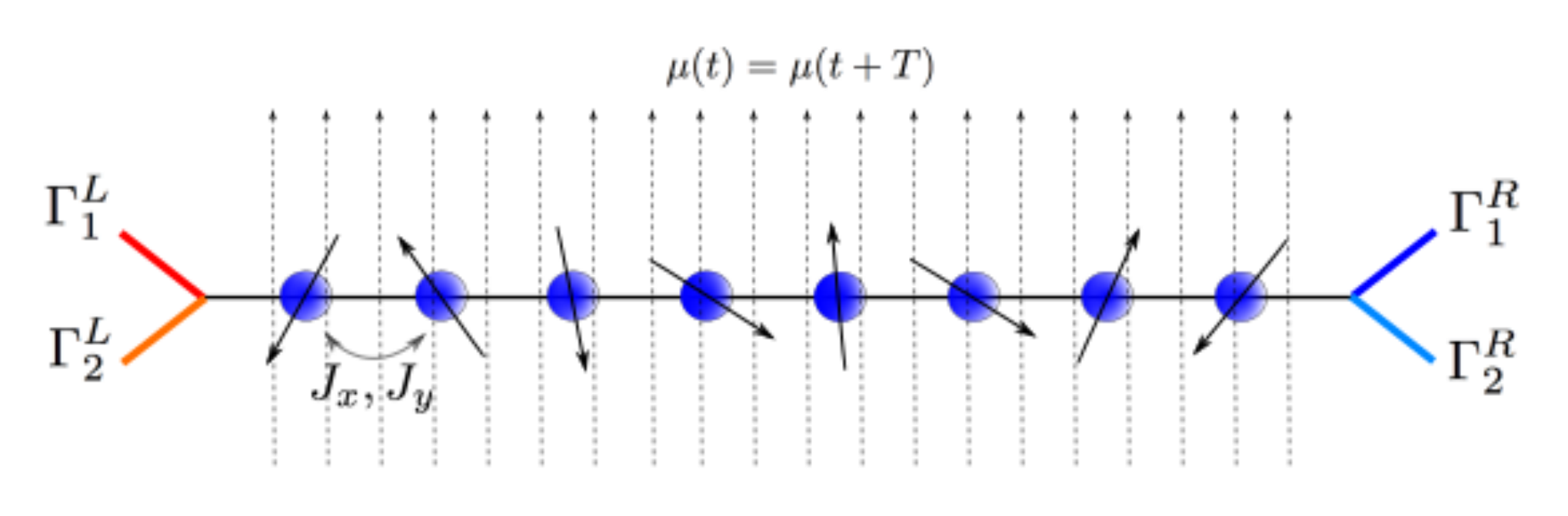}
		\caption{Schematic illustration of the driven spin-$\frac12$  chain coupled to dissipative baths at the ends.}
		\label{toy-model-pic}
	\end{figure}
\end{center}
\textit{Model} --- We consider  a spin-$\frac{1}{2}$ chain described by the  XY model  in a periodically driven transverse magnetic field. The system's Hamiltonian reads
\begin{align}
	\label{Hamiltonian_Ising}
	\mathcal{H}(t) = &- \sum_{n}  \left[ J_x \sigma_n^x \sigma_{n+1}^x + J_y \sigma_n^y \sigma_{n+1}^y + \mu(t) \sigma_n^z \right],
\end{align}
where $\sigma_n^a$ with $a = \left\{x,y,z\right\}$ is the Pauli matrix at site $n$. 
The exchange couplings are parametrized as $J_x = \frac{\gamma - \Delta}{2}$ and $J_y = \frac{\gamma + \Delta}{2}$ and $\mu$ is the time dependent transverse magnetic field. 
Via a Jordan-Wigner transformation~\cite{Jordan-Wigner}, the spin chain Hamiltonian~\eqref{Hamiltonian_Ising} can be mapped onto that of a fermionic model describing a $p$-wave
superconductor~\cite{Kitaev2001} or equivalently,  can be rewritten in terms of $2N$ Majorana fermions $w_i$   
\begin{align}
	\label{Hamiltonian_Majorana}
	\mathcal{H}(t) &= i \sum_{n=1}^{N-1} \bigg[ \frac{\gamma - \Delta}{2} w_{2n} w_{2n+1}  - \frac{\gamma + \Delta}{2}  	w_{2n-1} w_{2n+2} \bigg] \nonumber \\
	& \quad + i \sum_{n=1}^{N} \mu(t) w_{2n-1} w_{2n} = i \sum_{m,n}^{2N} w_m A_{mn}(t) w_n, 
\end{align}
where the Majorana operators $w_i$ satisfy the anti-commutation relations $\left\{w_i, w_j \right\} = 2 \delta_{ij}$.  
In the absence of  driving, the  closed XY-model in a field exhibits three distinct phases, two of which have nontrivial and opposite topology. 
Specifically,  in the topologically nontrivial phase for  $|\frac{\mu}{\gamma}| < 1$, zero-energy Majoranas appear at the ends of the chain~\cite{Thakurathi,Kitaev2001}. 
For the case of the  transverse field Ising model,  corresponding to $\gamma=\Delta$, periodic modulations of the transverse field were recently shown to  induce a multitude of FMF's for a wide range of system parameters, even when the undriven phase has  trivial topology~\cite{Thakurathi}. 
This is analogous to the generation of Floquet topological insulators from non-topological band insulators \cite{Katan-Podolsky,Lindner}.
 
We consider the case of  delta-function driving, modeled as $\mu(t)=\mu_0 + \mu_1 \sum_{n\in\mathbb{Z}} \delta(t-n T)$ where $T$ is the period of the drive.
The ensuing Floquet time-evolution operator over one period in the Majorana basis is given by a time-ordered exponential $U(T,0) = \mathcal{T} \left[ \exp \left( -i 
 \int_{0}^{T} \mathcal{H}(t) \mathrm{d}t \right) \right]$. 
By assuming periodic boundary conditions or an infinite chain, the Floquet operator then decouples into two-dimensional matrices described by the quasi-momentum $k$ ~\cite{Sedrakyan:2011,Thakurathi,DeGottardi1}:
\begin{equation}
U_k (T,0) =  e^{i \mu_1 \sigma^z}  e^{-i2T[(\gamma \cos k - \mu_0 ) \sigma^z + \Delta \sin k \sigma^y]} e^{i \mu_1 \sigma^z}
\label{Floquet-operator-delta}
\end{equation}
Typically, the number of generated FMF's is obtained by a direct evaluation of the topological winding number $W=\frac{1}{2\pi} \int_{BZ} \mathrm{d} \phi_k$, where $\phi_k=\tan^{-1}(a_{3,k}/a_{2,k})$ is an angle function derived from the effective Floquet Hamiltonian $h_{\mathrm{eff},k} = a_{2,k} \tau^y + a_{3,k} \tau^z \equiv i \log U_k(T,0)$, with Pauli matrices $\tau^{y,z}$\cite{Thakurathi,supmat}.
Here, we show  that the generation of FMF's can easily be understood via an analysis of the stationary points of the underlying Floquet energy dispersion.  The  eigenvalues $e^{i \theta_k}$ of the Floquet operator $U_k$  can be compactly written as
\begin{align}
\cos \theta_k &= \cos(2\mu_1) \cos(T E_{k,0}) +  \nonumber \\
& \quad +  \sin(2\mu_1) \frac{2(\gamma \cos k - \mu_0)}{E_{k,0}} \sin(T E_{k,0})
\label{delta_func_driving}
\end{align}
with $E_{k,0} = 2 \sqrt{(\gamma \cos(k) - \mu_0)^2 + \Delta^2 \sin^2(k)}$. 
These results can easily be generalized to the case of multi-step driving.

The eigenvalue equation for $\theta_k$ can be understood as a counterpart of the equation for Floquet  quasi-energies  $\epsilon_\alpha(k)$, defined as the eigenvalues of the operator $\mathcal{H}(t) - i \hbar \partial_t$ in $k$-space (or equivalently the exponent of the time-periodic Floquet wave function).
At topological transitions, quasi-energy gap closings in $\epsilon_\alpha(k)$ translate into the appearance of non-trivial stationary points in $\theta_k$. 
We find that the number $p$ of stationary points $k^*$, defined as $ \left. \mathrm{d} \theta_k / \mathrm{d} k \right|_{k=k^*}=0$ in the interval  $0<k \le \pi$ directly yields the number of  FMF's.
For given values of $\Delta$ and $\mu_0$, mapping the number of stationary points as a function of  $(T, \mu_1)$  provides a complex phase diagram shown in Fig. 2a) and e) with each phase being characterized by its own number of FMF's.  
Consequently, different FMF sectors are linked by a topological phases transition of the Lifshitz kind~\cite{Lifshitz}. 

These dynamical phase diagrams clearly illustrate that Floquet systems exhibit a rich and varied  $\mathbb{Z}_n$ topology that has no counterparts in the undriven system. 
However, any study of FMF's in an experimental context requires taking into account dissipation and also identifying accessible physical observables. 
To  this end,  we couple the chain to  two Markovian baths at the ends (see figure \ref{toy-model-pic}). 
For a weak coupling between the chain and the two reservoirs, the system realizes a non-equilibrium steady state (NESS).
In this weak coupling limit, we expect the time-evolution of the system's density matrix $\rho$ to be governed by the master equation in Lindblad form
\begin{equation}
	\dot{\rho}(t) = - i \left[ \mathcal{H}(t), \rho \right] + \hat{\mathcal{D}}_L(t)[\rho] + \hat{\mathcal{D}}_R(t)[\rho],
	\label{Lindblad-master-equation}
\end{equation}
where $\mathcal{H}(t)$ is the periodically driven Hamiltonian of Eq.~\ref{Hamiltonian_Majorana}.   
The relevant   dissipators  $\hat{\mathcal{D}}_L(t) = \sum_{\mu=1, 2} ( 2 L_{\mu} \rho L_{\mu}^{\dagger} - \left\{ L^{\dagger}_{\mu} L_{\mu}, \rho \right\} )$   and  $ \hat{\mathcal{D}}_R(t)= \sum_{\mu=3, 4} ( 2 L_{\mu} \rho L_{\mu}^{\dagger} - \left\{ L^{\dagger}_{\mu} L_{\mu}, \rho \right\} )$     
describe the effect of the coupling to the baths in terms of jump-operators $L_{1,2}(t) = \sqrt{\Gamma_{1,2}^L(t)} \left( w_1 \pm i w_2 \right)$ (left bath) and $L_{3,4}(t) = \pm (-i)^N \sqrt{\Gamma_{1, 2}^R(t)} \left( w_{2N-1} \pm i w_{2N} \right)$ (right bath).  The rates $\Gamma_{1,2}^{L,R}(t)$ completely characterize the effect of the bath on the system\cite{Prosen2011} and are in principle time-dependent. In the limit of weak dissipation and for the pulsed driving studied we can apply the time convolutionless approximation for periodically driven systems~\cite{Saeki:1986} and show that the assumption of time independent rates for our model is qualitatively justified, particularly for finite sized systems. Furthermore, the explicit inclusion of time periodic rates in the form of delta kicks does not lead to any significant deviation from the results using constant coefficients. Consequently, this choice of bath imposes a net incoherent magnetization of the end spins along the $z$ direction in the absence of any spin-spin interactions.  We mention that the directions of this  imposed magnetization can generally  lead to very different  scenarios  for certain observables in the steady state.

\begin{widetext}

\begin{figure}[H]
\centering
\includegraphics[width=\columnwidth]{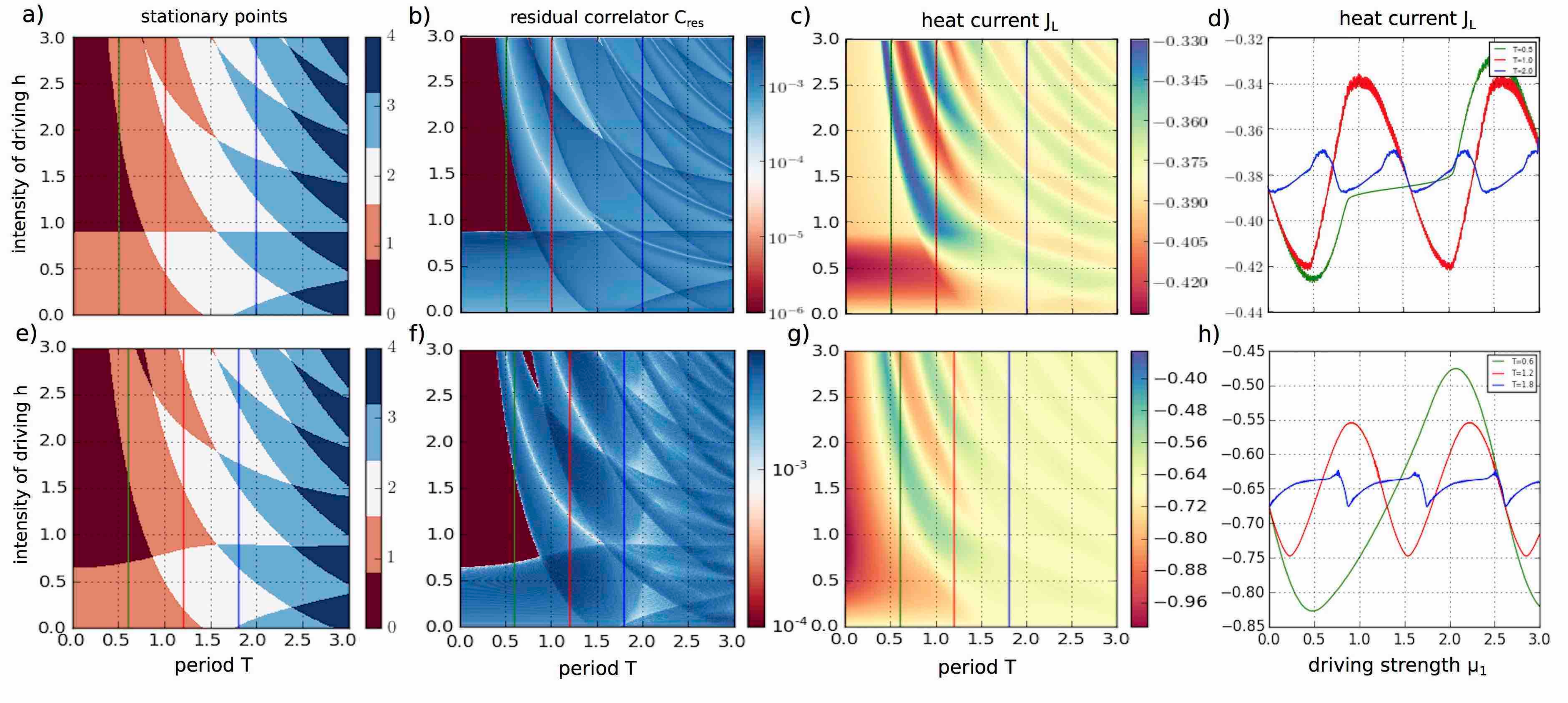}
\caption{Stationary point phase diagram of the Floquet operator (left-most  column), residual correlator $C_{res}$ (second column from the left) and heat current $J_L$ (third and fourth columns from the left) for two different starting points in the $(\mu_0,\Delta)$ plane. The residual correlator $C_{res}$ maps out the phase diagram obtained with the stationary points\cite{Prosen2011}. Figures 2a)-2d) correspond to the quasi-isotropic case in a low static field ($\Delta=0.1$, $\mu_0=0.1$), while  2e)-2g) show the results for moderate anisotropy ($\Delta=0.5$, $\mu_0=0.1$). The cuts displayed in 2d) and 2h) are indicated by vertical lines of corresponding color in Figs.2d) and 2h).}
\label{heat-current-plot}
\end{figure}

\end{widetext}

Rewriting the dissipators described above in the Majorana representation\cite{Prosen2011,supmat}, we find that equation \eqref{Lindblad-master-equation} for the time evolution of the density matrix can be recast as an equation for the covariance  matrix $C_{ij}(t) \equiv \Tr[w_i w_j \rho(t)] - \delta_{ij}$.  
The covariance matrix $C$ satisfies,
\begin{equation}\label{eq:covariance}
\dot{C}_{ij}(t) = -i \mathrm{Tr}\left[w_i w_j \left[\mathcal{H}(t), \rho(t)\right] \right] + \mathrm{Tr}\left[ w_i w_j \hat{\mathcal{D}}(\rho) \right],
\end{equation} 
where the r.h.s can be evaluated using Wick's theorem since both $\mathcal{H}$ and $\hat{\mathcal{D}}$ are quadratic in Majorana fermions.  

Although the full time dependent equation is not easy to solve, the stroboscopic behavior of the covariance matrix  in the steady state can be obtained using Floquet theory. 
Firstly, in the steady state, the stroboscopic covariance matrix $C_F = C(0) = C(T)$ will no longer depend on the initial conditions and will fully exhibit the periodicity of the underlying drive.
Following the treatment of Ref~\cite{Prosen2011}, the steady state behavior of the covariance matrix can be shown to be governed by the \textit{discrete Lyapunov equation}~\cite{Zhou}
\begin{equation}
	Q(T) C_F - C_F Q^{-T}(T) = iP(T),
	\label{discrete-lyapunov-cov-matrix}
\end{equation}
where the matrices $Q$ and $P$ depend on the nature of the driving\cite{supmat}. 
Solving this Lyapunov equation then helps us obtain various stroboscopic observables as a function of $C_F$.

In the absence of an order parameter to track the  topological phase transitions in our spin chains,  a weighted sum of the covariance matrix called the  \textit{residual correlator} $C_{res} \propto \sum_{|j-k| \ge N/2} |C_{j,k}|$ has been shown to play the role of an effective order parameter which tracks the stationary point phase diagram\cite{Prosen2011}. 
The structure of $C_{res}$  was already shown to give a one-to-one correspondence to the stationary points phase diagram in Ref. \cite{Prosen2011}. Ref. \cite{Prosen2011} however, did not relate this result to the topological nature of the transitions nor to the generation of FMFs.
Figs. 2b) and f) show that  the underlying stationary point phase diagram seen in the closed system survives even in the presence of dissipation, though the phase boundaries are mildly shifted. 
Unfortunately, although $C_{res}$ indicates the boundaries delineating regions with differing FMF's, it does not indicate the number of FMF's in each zone nor is it an easily accessible experimental observable. 
A natural observable would be stroboscopic spin correlation functions, which can be easily obtained from $C_F$.
However, these are not good trackers of the phase transitions as the associated signatures are weak. 
Charge transport, on the other hand has caveats as  highlighted in the introduction. 

We now show that a good candidate to probe the hierarchy of topological phase transitions and obtain  observable signatures related to the number of FMF's is the heat transport across the chain in the non-equilibrium steady state (NESS). 
 The heat currents from and to the reservoirs can be obtained from the first law of thermodynamics $\mathrm{d}U = \delta Q + \delta W$, where $\delta Q$ is the change in heat and $\delta W$ is the change in work. 
 The rate of change of the internal energy is given in terms of $\rho$ is given by
\begin{equation}
\frac{\mathrm{d} U}{\mathrm{d} t}=\mathrm{Tr}\left[ \dot{\mathcal{H}}(t) \rho(t) \right] + \mathrm{Tr}\left[ \mathcal{H} \dot{\rho}(t) \right],
\end{equation}
The first term is related to the power of the system, while the second corresponds to the change in heat.  
The Lindblad  equation for the density matrix \eqref{Lindblad-master-equation}  leads to the  following definition of the heat current:  $J_{L,R} \equiv  \Tr \left[ \hat{\mathcal{D}}_{L,R}(\rho) \mathcal{H} \right]$~\cite{Solinas}. 
Note that the direction of the current is implicitly contained in $J_L$ and $J_R$ by defining the quantities as the flow of heat from the reservoirs \emph{into} the system. 
Using the Majorana basis and  \eqref{eq:covariance}, the stroboscopic heat current of the left reservoir can be simply expressed in terms of the covariance matrix
\begin{align}
J_L &=  4 \Gamma_+^L \left[ i J_x C_{3,2} - i J_y C_{4,1} + 2i \mu_0 C_{2,1} \right] + 8\mu_0 \Gamma_-^L \label{eq:hcL} \\
J_R &=  4 \Gamma_+^R \bigg[ i J_x C_{2N-1,2N-2} - i J_y C_{2N,2N-3} + \nonumber \\
& \quad + 2i \mu_0 C_{2N,2N-1} \bigg] + 8\mu_0 \Gamma_-^R. \label{eq:hcR}
\end{align}
Our results for the stroboscopic heat current $J_L$ are shown in Fig. 2 for different driving parameters and weak static field $\mu_0$.
In Figs. 2a) and e), the corresponding Lifshitz points phase diagrams indicating the number of FMF's are plotted.
Figs. 2b), 2f) resp. 2c), 2g) show the behaviours of $C_{res}$  resp. $J_L$ across the phase diagram. 
Clearly, $C_{res}$ delineates the different topological phases even in the presence of dissipation. Reversals of the heat current  flow in topological phases with nonzero FMF's  are clearly seen. This  is essentially due to driving and 
 depends on   the  choice of bath parameters $\Gamma_{1,2}^{L,R}$  within a FMF phase.
 To see specific features of the heat current  at these topological phase boundaries we consider the vertical cuts plotted in Figs.~2d) and h). 
  Typically,  amplitudes of the heat current decreases as the number of FMF's increases.
We find that, at the phase boundaries between two phases with differing  non-trivial topology, the slope of the heat current, with respect to the tuning parameter, changes sign. 
On the other hand, transitions between a zero and a nonzero FMF phase are tracked by either changes in sign or discontinuities in the slope of the heat current.

Away from the transitions, the quasi-energy spectra is completely gapped and the heat transport is  essentially mediated by FMF's.  
The high-frequency oscillations in the heat current are due to finite size effects  and decrease with increasing $N$.
 This change in sign of the slope of the heat current with respect to the control parameter effectively tracks the parity of the phase and  is valid for any cut in the phase diagram. 
Since the actual sign of the heat current is determined by the bath parameters, it is not possible to assign a fixed parity to a phase, rather heat current is sensitive only to changes in parity.
Consequently, one cannot  ascertain whether a given phase has even or odd number of FMF's.  
For certain bath parameters, the heat current can indeed change sign within a given topological phase without a concomitant change in the sign of the slope of the current.
 
Analogous conclusions can be drawn from the analysis of the heat current from the right reservoir or the net heat flow $J_L + J_R$.
It is important to note that the net heat flow is not necessarily zero, since the driving has the effect of injecting energy into the system, which can then preferentially extract or dump excess heat in one or the other reservoir, depending on the physical state of the chain and the details of the baths. 
To summarize, the heat current is a sensitive detector of the topological phase transitions for a wide
 range of static magnetic fields, provided the anisotropy $\Delta \leq 0.5$, whereas the signals lose their precision for high-anisotropy states.  
 
It is reasonable to expect other observables to  be sensitive to these
phase transitions.  An example in the
dissipative set up studied here is
the spin current.  Using the formalism described here, we find that though the spin current at the ends manifests changes at the
phase boundaries, these were found to be far too weak to provide the requisite smoking gun  evidence.  The heat current is far
more sensitive an observable.

\begin{figure}[H]
		\includegraphics[width=\columnwidth]{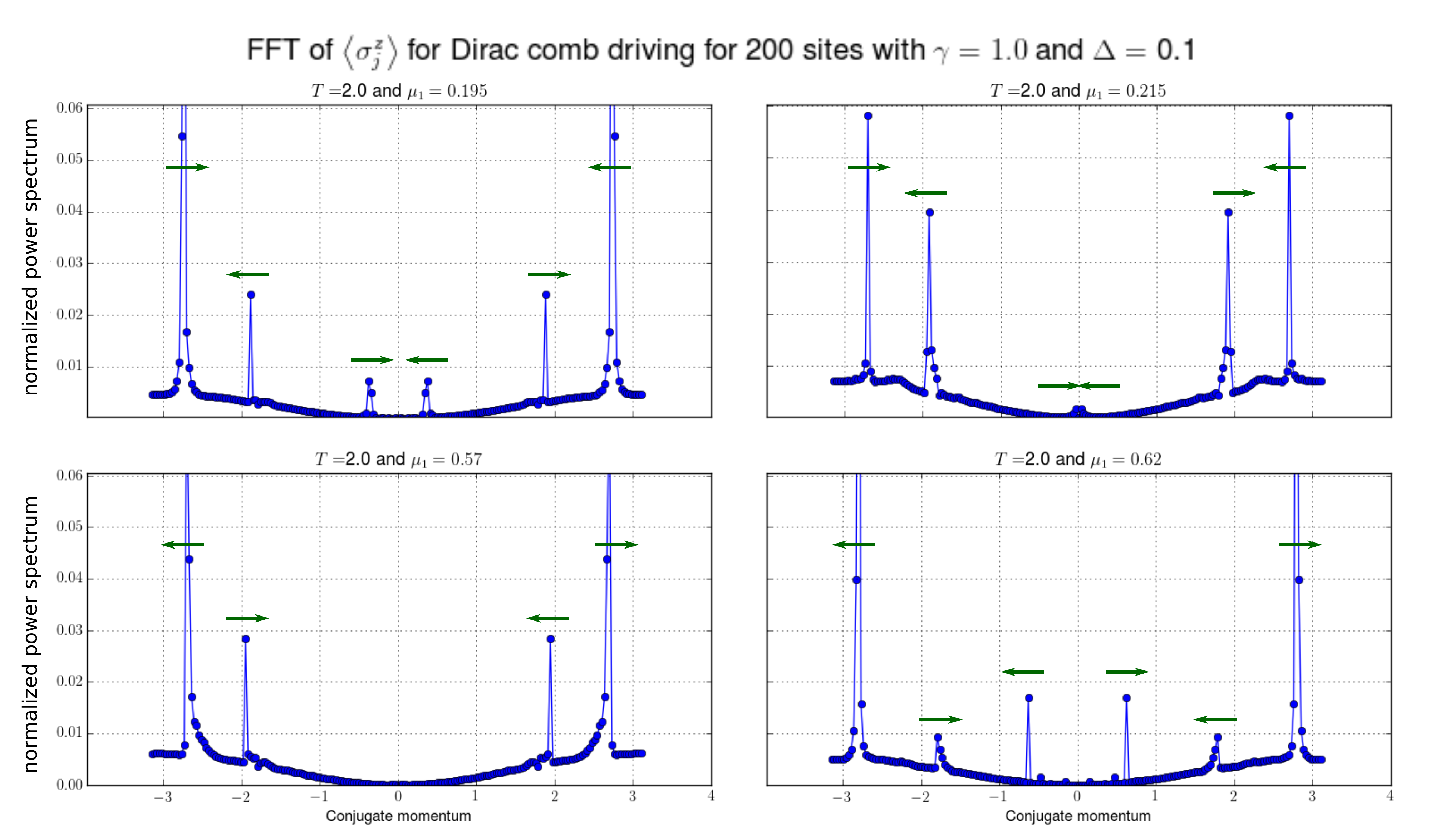}
		\caption{Power spectrum of the spin density profile for delta-kick driving with period $T=2.0$ at increasingly larger intensity $\mu_1$. The system was driven out of a quasi-isotropic regime with $\Delta = 0.1$ and $\mu_0=0.0$. The green arrows indicate in which direction the peaks shift upon increasing $\mu_1$. The total number of peaks is twice the number of FMF's.}
		\label{spin-profile-plot}
\end{figure}

We  now show that the number of FMF's in any phase  can directly be read off
from the Fourier transform of the spatial spin profile $\left< \sigma_i^z \right>$.  
In the driven setup considered here, the spin profile  away from the edges is distinctly non-uniform in all the nontrivial Floquet topological phases.   
This is because the FMF's are not localized in the spin language. 
The Fourier transform of the spin profile is plotted in Fig.~\ref{spin-profile-plot} for several values of the driving intensity across the cut at period $T=1.0$. 
We see that the presence of FMF's is manifested by the appearance of  pronounced peaks which correspond to superpositions of the modulations of the spin profile at different $k$ vectors.  
Note that in  Fig.~\ref{spin-profile-plot}, the central peak corresponding to the uniform background has been removed to facilitate the visualization of the other peaks.

As the control parameter is varied, these symmetric pairs peaks move smoothly either move towards the origin or  away from it. 
The green arrows in figure~\ref{spin-profile-plot} indicate the direction of the movement as the intensity of the driving is increased. 
Peaks are destroyed or created at zero momentum, indicating that coherent spin modulations appear/disappear above a uniform background signaling the creation or annihilation of FMF's.  
The  total number of peaks  with $k \ne 0$, obeys $N= 2 p$ where $p$ is the number of FMF's in the given phase determined by the control parameters. 
As for the heat current, this counting of FMF from the spin density is robust for small anisotropies. 
At higher anisotropy ($\Delta > 0.5$) and higher periods ($T > 2.0$) the correspondence between localized peaks and FMF's loses its precision as new peaks appear at fixed momenta, possibly corresponding to other forms of excited many-body states.

We now discuss a plausible physical connection between heat current and the peaks  that appear in the  Fourier transform of the spin density profile. 
Typically a peak  in the Fourier transform  signals coherent modulations of the spin density  with a driving-dependent wavelength  which  can carry energy from one end of the chain to the other, with higher-wavelength modulations carrying more energy than lower-frequency ones. 
However, having more peaks does not automatically translate to increased energy transport as multiple FMF's can lead to destructive interference or standing waves in the spin profile modulations, reducing the ability of the system to carry heat.  
This feature is highlighted for instance at $T=2.0$ and $\mu_1$ in figure \ref{spin-profile-plot}, where the two Fourier peaks resonate at the same wavelength and simultaneously the heat current is zero.  

\textit{Conclusions} --- 
In conclusion, we have shown that the hierarchy of topological phase transitions generated in a periodically driven dissipative system can be easily detected via heat transfer.
The heat/energy current that flows through the chain in its NESS  tracks the series of phase transitions generated when Floquet Majorana modes are created or destroyed.  
Furthermore,  the spin density profile  provides a simple way of counting the number of  Floquet modes  present in a phase.
Direct detection of exotic Majorana modes is consequently easier, as the ability to tune through the cascade of phase transitions  automatically eliminates the question of distinguishing other modes --- like Andreev bound states --- which mimic the Majorana modes. 
Furthermore, spin chains can be simulated in quantum engineered systems either using trapped ions\cite{Monroe} or flux qubits\cite{Schoen}, where many such units can be combined to realize potentially long chains. 
Both systems offer a controllable ways to apply periodic magnetic fields.  
The FMF counting scheme can be easily implemented in trapped ions using single site fluorescence. 
The switching of the stroboscopic heat current  seen in this system opens up the intriguing possibility of using  Floquet Majorana phases to devise both quantum heat engines or heat pumps.  
To establish such topology driven functionality, more in-depth studies  of the work done during a time period as well as the influence of time dependent dissipative coefficients are required, which is beyond the scope of the present paper.
Future directions involve the study  of interactions and the role played by dephasing of the spins on the robustness of the FMF's discussed here as well as signatures specific to such dynamically induced topological phase transitions. It would be interesting to study how the richness of the phenomena seen in this simple one dimensional chain generalize to models in higher dimensions.

\textit{Acknowledgments}
This project is supported by SNF, Mr. Giulio Anderheggen and the ETH Z\"{u}rich Foundation.


\begin{thebibliography}{99}

\bibitem{Eisert} J. Eisert and M. Friesdorf and C. Gogolin, Nature Physics 11, 124Ð130 (2015).
\bibitem{Lazarides} A. Lazarides, A. Das, and R. Moessner, Phys. Rev. E 90, 012110, (2014).
\bibitem{Eckardt} A. Eckardt, C. Weiss, and M. Holthaus, Phys. Rev. Lett. \textbf{95}, 260404 (2005).
  \bibitem{Grifoni} M. Grifoni and P. H\"{a}nggi, Phys. Rep. \textbf{304}, 229 (1998).
  \bibitem{Haenggi} T. Dittrich, P. H\"{a}nggi, G.-L. Ingold, B. Kramer, G. Sch\"{o}n, and W. Zwerger,, \textit{Quantum Transport and Dissipation}, Wiley-VCH, 1998.
\bibitem{OdedChitra} R. Chitra and O. Zilberberg, Phys. Rev. A 92, 023815 (2015).
\bibitem{Lindner} N. H. Lindner, G. Refael, and V. Galitski, Nature Phys. 7, 490 (2011).
\bibitem{Kitagawa} T. Kitagawa, E. Berg, M. Rudner and E. Demler, Phys. Rev. B 82, 235114 (2010).
\bibitem{Titum} P. Titum, N. H. Lindner, M. C. Rechtsman and G. Refael, Phys. Rev. Lett. 114, 056801 (2015).
\bibitem{Gedik} Y. H. Wang, H. Steinberg, P. Jarillo-Herrero and N. Gedik, Science 342, 453 (2013).
 \bibitem{Kitaev2001} A. Yu. Kitaev, Usp. Fiz. Nauk. \textbf{171} (10), 131 (2001).
\bibitem{Kitaev2} A. Y. Kitaev, Ann. Phys. 302, 2 (2003).
 \bibitem{Liu} D. E. Liu, A. Levchenko and H. U. Baranger, Phys. Rev. Lett. \textbf{111}, 047002 (2013).
\bibitem{AliceaReview} J. Alicea, Rep. Prog. Phys. \textbf{75} 076501 (2012).
\bibitem{KTLaw} K. T. Law, P. A. Lee and T. K. Ng, Phys. Rev. Lett. \textbf{103}, 237001 (2009).
\bibitem{Fidkowski} L. Fidkowski, J. Alicea, N. H. Lindner, R. M. Lutchyn, and M. P. A. Fisher, Phys. Rev. B \textbf{85}, 245121 (2012).
\bibitem{Fu-Kane2009} L. Fu and C. L. Kane, Phys. Rev. Lett. \textbf{102}, 216403 (2009).
\bibitem{Mourik} V. Mourik, K. Zuo, S. M. Frolov, S. R. Plissard, E. P. A. M. Bakkers, and L. P. Kouwenhoven, Science \textbf{336}, 1003 (2012).
\bibitem{Woerkom:2016aq} D. J. van Woerkom, A. Proutski, B. van Heck, D. Bouman, J. I. V\"{a}yrynen, L. I. Glazman, P. Krogstrup, J. Nyg?ard, L. P. Kouwenhoven, and A. Geresdi (2016), 1609.00333, URL \url{https://arxiv.org/abs/1609.00333}.
\bibitem{Plugge:2016cy} S. Plugge, A. Zazunov, E. Eriksson, A. M. Tsvelik, and R. Egger (2016), 1601.04332, URL \url{https://arxiv.org/ abs/1601.04332}.
\bibitem{Sedrakyan:2011} T. A. Sedrakyan and V. M. Galitski, Phys. Rev. B 83, 134303 (2011).
\bibitem{Thakurathi} M. Thakurathi, A. A. Patel, D. Sen and A. Dutta, Phys. Rev. B \textbf{88}, 155133 (2013).
\bibitem{Kundu} A. Kundu and B. Seradjeh, Phys. Rev. Lett. \textbf{111}, 136402 (2013).
\bibitem{YantaoLi} Y. Li, A. Kundu, F. Zhong, and B. Seradjeh, Phys. Rev. B \textbf{90}, 121401(R) (2014).
\bibitem{Farrell-Pereg-Barnea} A. Farrell and T. Pereg-Barnea, Phys. Rev. B \textbf{93}, 045121 (2016).
 \bibitem{Jordan-Wigner} P. Jordan and E. Wigner, Zeitschrift f\"{u}r Physik \textbf{47}, No. 9. (1928).
\bibitem{Katan-Podolsky} Y. T. Katan and D. Podolsky, Phys. Rev. Lett. \textbf{110}, 016802 (2013).
  \bibitem{DeGottardi1} W. DeGottardi, D. Sen, and S. Vishveshwara, New. J. Phys \textbf{13}, 065028 (2011).
\bibitem{Lifshitz} I. M. Lifshitz, Sov. Phys. JETP 11 N. 5, 1130 (1960).
\bibitem{Prosen2011} T. Prosen and E. Ilievski, Phys. Rev. Lett. \textbf{107}, 060403 (2011).
\bibitem{Saeki:1986} M. Saeki, J. Phys. Soc. Jap. 55 No. 6, 1861 (1986).
\bibitem{supmat} \textit{Supplementary material}.
  \bibitem{Zhou} K. Zhou, J.C. Doyle and K. Glover, \textit{Robust and Optimal Control} (Prentice Hall, New Jersey, 1995).
\bibitem{Solinas} P. Solinas, D. V. Averin and J. P. Pekola, Phys. Rev. B 87, 060508(R), (2013).
\bibitem{Monroe} E. E. Edwards, S. Korenblit, K. Kim, R. Islam, M.-S. Chang, J. K. Freericks, G.-D. Lin, L.-M. Duan and C. Monroe, Phys. Rev. Lett. B 82, 060412(R) (2010).
\bibitem{Schoen} Y. Makhlin, G. Sch\"{o}n and A. Shnirman, Rev. Mod. Phys. 73, 357 (2001).
\bibitem{Prosen2008} T. Prosen and I. Pi\vaccent{z}orn, Phys. Rev. Lett. \textbf{101}, 105701 (2008).
\bibitem{Prosen2008NJP} T. Prosen, New J.  Phys. \textbf{10}, 043026 (2008).






\end{thebibliography}
\end{document}